\documentclass[aip,jcp,10pt,twocolumn,superscriptaddress,nofootinbib,reprint]{revtex4-1}

\usepackage{bm}
\usepackage[english]{babel}
\usepackage{graphicx}
\usepackage{hyperref}
\usepackage[utf8]{inputenc}
\usepackage{mathtools}
\usepackage[expansion=true,protrusion=true]{microtype}
\usepackage{siunitx}
\usepackage{textcomp}
\usepackage[usenames,dvipsnames]{xcolor}

\hypersetup{hidelinks}

\usepackage{cleveref}

\renewcommand{\eqref}[1]{(Eq.~\ref{#1})}
\crefformat{equation}{(Eq.~#2#1#3)}

\newcommand*{\citen}[1]{%
  \begingroup
    \romannumeral-`\x 
    \setcitestyle{numbers}%
    \cite{#1}%
  \endgroup   
}

\newcommand{\U}[1]{\,\mathrm{#1}}

\newcommand{\lB}{l_{B}}
\newcommand{\lD}{\lambda_{D}}
\newcommand{\kT}{k_{B}T}
\newcommand{\tens}[1]{\bar{\bm{#1}}}
\let\vec\bm

\newcommand{\ie}{i.e.,~}

\usepackage{natbib}

\begin{document}

\title{Limiting Spurious Flow in Simulations of Electrokinetic Phenomena}

\author{Georg Rempfer}
\email{georg.rempfer@icp.uni-stuttgart.de}
\author{Gary B. Davies}
\author{Christian Holm}
\affiliation{Institute for Computational Physics (ICP), University of Stuttgart, Allmandring 3, 70569 Stuttgart, Germany}

\author{Joost de Graaf}
\email{j.degraaf@ed.ac.uk}
\affiliation{School of Physics and Astronomy, University of Edinburgh, Scotland, Edinburgh EH9 3JL, United Kingdom.}

\date{\today}

\begin{abstract}

Electrokinetic transport phenomena can strongly influence the behaviour of macromolecules and colloidal particles in solution, with applications in, e.g., DNA translocation through nanopores, electro-osmotic flow in nanocapillaries, and electrophoresis of charged macromolecules. Numerical simulations are an important tool to investigate these electrokinetic phenomena, but are often plagued by spurious fluxes and spurious flows that can easily exceed physical fluxes and flows. Here, we present a method that reduces one of these spurious currents, spurious flow, by several orders of magnitude. We demonstrate the effectiveness and generality of our method for both electrokinetic lattice-Boltzmann and finite-element-method based algorithms by simulating a charged sphere in an electrolyte solution, and flow through a nanopore. We also show that previous attempts to suppress these spurious currents introduce other sources of error.

\end{abstract}

\maketitle

\section*{Introduction}
\label{sec:intro}

Electrokinetic transport phenomena play a major role in the dynamics of macromolecules and colloidal particles in solution. The intricate interplay between diffusion, electrostatics, and hydrodynamics makes them an interesting topic for theoretical research. Applications based on electrokinetics comprise, among others, the characterization and separation of biomolecules, colloids and ions,~\cite{obrien_electrophoretic_1978, ohshima_electrophoresis_1995, ohshima_electrophoretic_1995, hill03a, smeets_salt_2006, dekker_solid-state_2007, dhopeshwarkar_transient_2008, lan_nanoparticle_2011, german_controlling_2013, lan_effect_2014} microfluidic pumping and rectification mechanisms,~\cite{burgreen_electrokinetic_1964, rice_electrokinetic_1965, daiguji_nanofluidic_2005, white_ion_2008, berg_exact_2009, laohakunakorn_electroosmotic_2015, laohakunakorn_electroosmotic_2015-1} active particles,~\cite{de_graaf_diffusiophoretic_2015, brown14, moran10, moran11, sabass12b, kreissl, brown15a-pre} and model systems for living cells.~\cite{hsu_diffusiophoresis_2010} Gaining experimental understanding of each of these mechanisms can be difficult due to the inherently small length scales involved. Therefore, theoretical and numerical studies are often used to complement and provide insight to experimental work. 

The electrokinetic equations can be solved analytically for a number of special cases, giving insight into phenomena such as electric double layers and electrostatic screening,~\cite{huckel1923electrolytes} and electrophoresis in the limit of high~\cite{smoluchowski_contribution, henry_cataphoresis_1931} and low~\cite{huckel1924kataphorese} salt concentrations. Analytical techniques based on first order perturbation expansions of the applied electric field have been used to determine the electrophoretic mobility of bare colloids and colloids homogeneously grafted with polymers and polyelectrolytes.~\cite{obrien_electrophoretic_1978, hill03a, ohshima_electrophoresis_1995, ohshima_electrophoretic_1995} 

Electrokinetic phenomena can also be simulated using mesoscopic methods such as dissipative particle dynamics,~\cite{groot_electrostatic_2003,smiatek_mesoscopic_2011} multi-particle collision dynamics,~\cite{frank08a,frank09a,gompper08a} or hybrid particle-lattice-Boltzmann algorithms.~\cite{ahlrichs99a, lobaskin04b,lobaskin07a,grass08a,raafatnia14b,fahrenberger15b} See Refs.~[\citen{rotenberg_electrokinetics_2013}] and~[\citen{slater09a}] for a more thorough review of these different mesoscopic methods. 

However, these particle-based algorithms are generally restricted to systems on the nanoscale, and analytical techniques based on perturbation theory cannot model non-linear effects caused by large applied electric fields; for systems on experimental length scales with few symmetries, the fully non-linear, coupled equations must be solved numerically. In recent years, growing computational power has allowed researchers to simulate electrokinetic phenomena on experimentally relevant length and time scales using continuum simulations of the electrokinetic equations.~\cite{white_ion_2008, laohakunakorn_electroosmotic_2015, laohakunakorn_electroosmotic_2015-1,laohakunakorn_dna_2013,  german_controlling_2013, lan_effect_2014, lan_nanoparticle_2011}

One popular method to solve the continuum electrokinetic equations is the finite-element method, a technique that involves reformulating the equation in the so-called weak formulation and using a finite set of basis functions to derive a discrete equation system that can be solved numerically. Finite-element simulations of the electrokinetic equations have been used to investigate electrophoresis of macromolecules, electro-osmotic flow in nanopores,~\cite{white_ion_2008, laohakunakorn_electroosmotic_2015, laohakunakorn_electroosmotic_2015-1} and translocation of colloids~\cite{lan_nanoparticle_2011, german_controlling_2013, lan_effect_2014} and DNA.~\cite{laohakunakorn_dna_2013}

Capuani~\textit{et al.}\cite{capuani_discrete_2004} developed an alternative to the finite-element method by devising a numerical algorithm that combines the finite-volume, finite-difference, and lattice-Boltzmann method.~\cite{capuani_discrete_2004} This algorithm was later used to investigate the electrophoretic mobility of charged spherical particles as a function of the Pecl\'et number.~\cite{giupponi_colloid_2011}

We show in this article that extreme care must be taken when discretising the electrokinetic equations, and that if the conventional form of the electrokinetic equations based on the available literature~\cite{rice_electrokinetic_1965, obrien_electrophoretic_1978, burgreen_electrokinetic_1964, capuani_discrete_2004, giupponi_colloid_2011, white_ion_2008, berg_exact_2009, daiguji_nanofluidic_2005, dhopeshwarkar_transient_2008} is used, significant errors arise. In the best case scenario, these errors can be controlled by using highly refined grids, requiring significant computational effort. In the worst case scenario, they lead to incorrect and unphysical results. 

Here, we present a new model that reduces errors due to spurious flow at constant computational cost by incorporating an additional gradient term into the hydrodynamic part of the continuum electrokinetic equations. While the problems and solutions we discuss are relevant to any discretisation method or non-equilibrium electrokinetic phenomenon, we illustrate the issues and improvements using two commonly used schemes from the literature: a numerical solver for the time-dependent electrokinetic equations by Capuani \textit{et al.}\cite{capuani_discrete_2004}\  and a solver for the stationary electrokinetic equations based on the finite-element method.~\cite{white_ion_2008}

We carry out simulations of a stationary charged sphere in an electrolyte solution using both methods. We show firstly the presence of spurious flow even in equilibrium, and then the corresponding reduction of spurious flow by several orders of magnitude using our improved model that incorporates a gradient term into the hydrodynamic part of the electrokinetic equations. Finally, we simulate a charged nanopore system and show that using our proposed gradient term reproduces the correct physics, where as methods based on the current literature do not. 

This article is organised as follows. \Cref{sec:governing_equations} reviews the physics of the continuum electrokinetic equations. In \Cref{sec:numerics}, we introduce the two most commonly used numerical schemes to solve these continuum electrokinetic equations, and explain the origin of the spurious fluxes and flows that occur in numerical simulations of electrokinetic phenomena. \Cref{sec:methods} describes the simulation setup and parameters that we used to simulate the stationary charged sphere system, and we present the results of these simulations in \Cref{sec:results}. \Cref{sec:conclusion} concludes the article.

\section{The Governing Equations for Electrokinetic Transport}
\label{sec:governing_equations}

The electrokinetic equations model the motion of charged species by a diffusion-advection equation. The expression for the flux of the charged species reads:
\begin{equation}
\label{eq:model_fluxes}
\vec{j}_{k} = \underbrace{-D_{k} \nabla c_{k} - \mu_{k} z_{k} e c_{k} \nabla \Phi}_{\vec{j}_{k}^{\mathrm{diff}}} + \underbrace{c_{k} \vec{u}}_{\vec{j}_{k}^{\mathrm{adv}}} ,
\end{equation}
where $c_{k}$ denotes the concentration of the ionic species with index $k$, $e$ denotes the elementary charge, and $\Phi$ the electric potential. The flux can be split into a diffusive, $\vec{j}_k^{\mathrm{diff}}$, and an advective, $\vec{j}_k^{\mathrm{adv}}$, part. \\
The diffusive flux itself has two contributions. The first term, often called the Fickian term, concerns the diffusion of ions due to local concentration variations: the ions flow from regions of high concentrations to regions of low concentration, and they do so with a diffusion coefficient $D_{k}$ that dictates the speed of diffusion.
The second contribution to the diffusive flux, sometimes called the migrative term, occurs due to the presence of local electric fields: ions with a charge $z_{k}e$ flow from regions of high potential to low potential, and they do so with a mobility $\mu_{k}$ that determines the speed of the ions' movement due to the applied electric field.

The diffusion coefficient and mobility are related by the Einstein-Smoluchowski relation $D_{k} / \mu_{k} = \kT$,~\cite{einstein_uber_1905,von_smoluchowski_zur_1906} where $k_B$ is Boltzmann's constant and $T$ the absolute temperature. \\

The advective flux, $\vec{j}_k^{\mathrm{adv}}$ accounts for the contribution of the underlying fluid's velocity $\vec{u}$ to the motion of the ionic species.

Without sources or sinks of the ionic species $k$, the diffusive flux of the $k^{\mathrm{th}}$ ionic species must fulfil the continuity equation:
\begin{equation}
\label{eq:model_continuity}
\partial_{t} c_{k} = -\nabla \cdot \vec{j}_{k} .
\end{equation}
For stationary situations in which none of the fields vary over time ($\partial_{t} = 0$), \cref{eq:model_fluxes} and \cref{eq:model_continuity} may be combined to form:
\begin{equation}
\label{eq:model_concentration}
\nabla \cdot (-D_{k} \nabla c_{k} - \mu_{k} z_{k} e c_{k} \nabla \Phi + c_{k} \vec{u}) = 0 .
\end{equation}

In order to solve~\cref{eq:model_concentration}, we need a model that describes both the electrostatic potential $\Phi$ and the fluid velocity $\vec{u}$. \\

Modelling the electrostatic potential is relatively straightforward: in the stationary state, and when the only source of electric field is due to the arrangement of the charges themselves, the electrostatic potential $\Phi$ fulfils the Poisson equation:
\begin{equation}
\label{eq:model_poisson}
\nabla \cdot (\varepsilon \nabla \Phi) = -\varrho = -\textstyle\sum_{k} z_{k} e c_{k} .
\end{equation}
Here the charge density $\varrho$ is given in terms of the ionic species concentrations $c_k$. The permittivity $\varepsilon = \varepsilon_{0} \varepsilon_r(\vec{r})$ is the product of the vacuum permittivity $\varepsilon_{0}$ and the local relative permittivity $\varepsilon_r(\vec{r})$ of the medium.

The hydrodynamics of soft matter systems usually concerns the motion of nano- to micrometer sized objects in fluids in which viscous forces dominate, \ie\ the low-Reynolds number regime. The motion of the fluid in this regime is governed by Stokes' equations:
\begin{equation}
\label{eq:model_stokes}
\begin{aligned}
\eta \bm{\nabla^{2}} \vec{u} &= \nabla p - \vec{f} ,\\
\nabla \cdot \vec{u} &= 0 .
\end{aligned}
\end{equation}
Here, $p$ denotes the hydrostatic pressure and $\eta$ the shear viscosity. The external body-force density $\vec{f}$ couples the fluid motion to the motion of the charged species. 
We note that even at concentrations of $1\U{mol/l}$, the charged species contribute at most a few percent to the mass of the total solution, which allows one to neglect the ionic species density in the hydrodynamic equations.~\cite{de_graaf_diffusiophoretic_2015,capuani_discrete_2004} \\

The fluid coupling body-force is commonly chosen as:~\cite{hsu_diffusiophoresis_2010, rice_electrokinetic_1965, burgreen_electrokinetic_1964, laohakunakorn_electroosmotic_2015, laohakunakorn_electroosmotic_2015-1, obrien_electrophoretic_1978, berg_exact_2009, white_ion_2008, daiguji_nanofluidic_2005, hlushkou_numerical_2005, dhopeshwarkar_transient_2008}
\begin{equation}
\label{eq:model_forcedens_estatics}
\vec{f} = \varrho \vec{E} = -\textstyle\sum_{k} z_{k} e c_{k} \nabla \Phi ,
\end{equation}
where $\vec{E}=-\nabla\Phi$ is the electric field.
In this case, the driving force $\vec{f}$ for the fluid is simply the net force acting on all the ionic species. In a stationary situation, there cannot be any momentum change in the ionic species: all momentum transported into the ionic species by the electric field must therefore be dissipated into the fluid. 

In situations where the ionic concentrations and the electrostatic field vary over time, momentum conservation for the ionic species does not hold. Yet, the force term~\cref{eq:model_forcedens_estatics} remains valid, since the time scale of the individual ions' acceleration is orders of magnitude smaller than the dynamics of the flow field and the ions' distributions.

To summarize, the time-dependent electrokinetic equations are given by the following system of equations
\begin{equation}
\begin{aligned}
\label{eq:model_summary}
\partial_{t} c_{k} &= \nabla \cdot (D_{k} \nabla c_{k} + \mu_{k} z_{k} e c_{k} \nabla \Phi - c_{k} \vec{u}) ,\\
\nabla \cdot (\varepsilon \nabla \Phi) &= {}-\textstyle\sum_{k} z_{k} e c_{k} ,\\
\eta \bm{\nabla^{2}} \vec{u} &= \nabla p + \textstyle\sum_{k} z_{k} e c_{k} \nabla \Phi ,\\
\nabla \cdot \vec{u} &= 0 .
\end{aligned}
\end{equation}
One obtains the stationary electrokinetic equations by setting the time derivative $\partial_{t} c_{k} = 0$.
This mean field model is valid in particular for moderate concentrations of monovalent ions without permanent magnetic moments in aqueous solution at room temperature, since it is based on the same principles as Poisson-Boltzmann theory.~\cite{andelman95a, holm01a}

\section{Numerical Solutions of the Electrokinetic Equations}
\label{sec:numerics}

In this section, we briefly discuss two common numerical schemes. In \Cref{sec:ek_solver}, we discuss a time-dependent solver that combines the finite-volume method (FVM), finite-difference method (FDM), and lattice-Boltzmann method (LBM), introduced by Capuani \textit{et al.}\cite{capuani_discrete_2004}; in \Cref{sec:fem_solver}, we discuss a solver that uses the finite-element method (FEM) to solve the complete set of stationary electrokinetic equations~\cref{eq:model_summary}. 

The time-dependent solver reproduces the stationary solutions by simulating sufficiently long times for the system to relax completely.

Solving the electrokinetic equations numerically introduces spurious fluxes and flows. In \Cref{sec:origin_spurious}, we discuss the origin of these numerical artefacts. 

\subsection{Solver Based on FVM, FDM, and LBM}
\label{sec:ek_solver}

Capuani \textit{et al.}\cite{capuani_discrete_2004}\  introduced a general method for solving the time-dependent electrokinetic equations~\cref{eq:model_summary} in their entire realm of applicability. They used this scheme to determine the electrophoretic mobility of spherical particles of various sizes in solutions of various salt concentrations.~\cite{giupponi_colloid_2011}

In a time-dependent simulation, it is essential to conserve the density of the ionic species to numerical precision: a drift in the net amount of ionic species would otherwise lead to a change in the net charge of the system and significantly influence the measured mobilities and conductivities.

For this reason, Capuani \textit{et al.}\cite{capuani_discrete_2004}\ propagate the densities according to the continuity equation ~\cref{eq:model_continuity} using a finite-volume scheme on a regular cubic grid. They calculate the discrete ionic fluxes between neighbouring grid nodes and the electrostatic potential using a simple finite-difference approximation of the diffusive flux expression in~\cref{eq:model_fluxes} and Poisson's equation~\cref{eq:model_poisson}. Instead of directly discretising the diffusive flux from~\cref{eq:model_fluxes}, they transform it into the following form:
%
\begin{equation}
\label{eq:flux_transformed}
 \resizebox{0.9\linewidth}{!}{$\vec{j}^\text{diff}_k = -D_{k} \exp\left(\frac{-z_{k}e\Phi(\vec{r})}{\kT}\right) \nabla \left[ c_{k}(\vec{r}) \exp\left(\frac{z_{k}e\Phi(\vec{r})}{\kT}\right) \right].$}
\end{equation}
Capuani \textit{et al.}\cite{capuani_discrete_2004}\  claim that this form of the expression suppresses spurious fluxes, since the gradient is applied to an approximately constant term if the concentrations are close to equilibrium, thus minimizing numerical errors. 

After applying a symmetric finite-difference discretisation over the link between two neighbouring nodes, this expression becomes
\begin{align}
\begin{split}
\label{eq:solver_flux}
& \resizebox{0.9\linewidth}{!}{$j^\text{diff}_{ki}(\vec{r}) = \frac{D_{k}}{2 \left| \vec{d}_{i} \right|}%
\left[ \exp\left(-\frac{z_{k}e\Phi(\vec{r})}{\kT}\right) + \exp\left(-\frac{z_{k}e\Phi(\vec{r} + \vec{d}_{i})}{\kT}\right) \right] \times$} \\
& \resizebox{0.9\linewidth}{!}{$\left[ c_{k}(\vec{r}) \exp\left(\frac{z_{k}e\Phi(\vec{r})}{\kT}\right) - c_{k}(\vec{r} + \vec{d}_{i}) \exp\left(\frac{z_{k}e\Phi(\vec{r} + \vec{d}_{i})}{\kT}\right) \right]$}
\end{split}
\end{align}
Here $j^\text{diff}_{ki}(\vec{r})$ denotes the flux of species $k$ from the node at position $\vec{r}$ to its neighbour at position $\vec{r} + \vec{d}_{i}$. 

The total fluxes are calculated according to~\cref{eq:solver_flux} and a volume of fluid scheme for the advective flux contribution in~\cref{eq:model_fluxes}. These fluxes can then be used to propagate the concentrations in time, according to the finite-volume representation of the continuity equation~\cref{eq:model_continuity}: 
\begin{equation}
c_{k}(\vec{r}, t + \Delta t) = c_{k}(\vec{r}, t) - \Delta t \textstyle\sum_{i}  A_{i} j_{ki}(\vec{r}, t) .
\end{equation}
Capuani \textit{et al.}\cite{capuani_discrete_2004}\  discretise the diffusive flux using face and edge neighbour nodes, and the advective flux using face, edge, and corner neighbours. This scheme maintains isotropy despite the underlying regular cubic grid, provided the weighting factors $A_{i} = A$. The remaining constant $A$ is chosen such that the discrete system reproduces the correct mean square displacement (MSD) for the ionic species in a situation of vanishing electric field and fluid velocity:
\begin{align}
\label{eq:msd_grid}
\textstyle\sum_{i} c(\vec{r} + \vec{d}_{i}, \Delta t) \vec{d}_{i}^2 = 6D\Delta t, \quad c(\vec{r}, 0) = \delta_{\vec{r}} .
\end{align}
This explicit propagation scheme is limited in stability. Using an implicit scheme would in principle allow for much larger time steps at the cost of having to solve a system of equations at every time step. However, since the LBM imposes a limit on the time step already, this advantage of an implicit scheme would be negated and the explicit scheme is the more sensible choice. 

To solve Poisson's equation~\cref{eq:model_poisson} for the electrostatic potential, they first assume that the electric permittivity $\varepsilon$ is constant, allowing one to express Poisson's equation~\cref{eq:model_poisson} as:
\begin{equation}
\label{eq:model_simple_poisson}
\nabla^2 \Phi = -4 \pi \lB \kT \varrho ,
\end{equation}
with the net charge density $\varrho = \sum_{k} z_{k}e c_{k}$ and the Bjerrum length $\lB = e^{2} / (4 \pi \varepsilon \kT)$. Discretising Poisson's equation~\cref{eq:model_simple_poisson} using a symmetric 7-point finite-difference stencil for the Laplacian, on the same regular cubic grid with grid spacing $h$ yields the following system of coupled linear equations:
\begin{align}
\label{eq:model_stencil}
\nonumber
{-4} \pi h^{2} \lB \kT \varrho(x,y,z) =~ & \Phi(x+h,y,z) + \Phi(x-h,y,z) \\\nonumber
+~&\Phi(x,y+h,z) + \Phi(x,y-h,z) \\\nonumber
+~&\Phi(x,y,z+h) + \Phi(x,y,z-h) \\
{-6} & \Phi(x,y,z) .
\end{align}

Capuani \textit{et al.}\cite{capuani_discrete_2004}\  solve this system using an iterative successive over-relaxation (SOR) scheme that benefits from the previous time step's solution as an initial guess. \\
To solve Stokes' equations~\cref{eq:model_stokes}, they employ a compressible single-relaxation-time lattice-Boltzmann scheme. \\
In~\Cref{sec:nonlinear_stencil_problem}, we show that the discretised flux expression~\cref{eq:solver_flux} proposed by Capuani \textit{et al.}\cite{capuani_discrete_2004}\  only reproduces the correct flux in special cases, and in general results in an error that increases exponentially with the potential gradient. We present a method that fixes this problem and eliminates the unbounded exponential error. 

\subsection{Solvers Based on the Finite-Element Method}
\label{sec:fem_solver}

Let us now turn our attention to the second common method of solving the EK equations.

In order to solve the electrokinetic equations using the finite-element method, one must reformulate the stationary electrokinetic equations~\cref{eq:model_summary} in the so-called weak form. In the weak formulation, we multiply both sides of the equations by a test function $\varphi$ and integrate over the whole domain, rather than requiring that the fields $c$, $\Phi$, $\vec{u}$, and $p$ satisfy the partial differential equations~\cref{eq:model_summary} directly (the strong form). As an example, carrying out this procedure for Poisson's equation yields:
\begin{align}
\label{eq:weak_form}
\int_\Omega \varphi \nabla^2 \Phi \, dV &= -\int_\Omega \varphi \varrho / \varepsilon \, dV \notag \\
\Leftrightarrow \int_\Omega \nabla \varphi \nabla \Phi \, dV &= \underbrace{\int_{\partial\Omega} \varphi \nabla \Phi \, d\vec{A}}_{= 0\text{, since } \varphi(\partial\Omega) = 0} + \int_\Omega \varphi \varrho / \varepsilon \, dV ,
\end{align}
using Green's first identity and assuming that the test function $\varphi$ vanishes on the domain boundary, and that $\varepsilon$ is spatially homogeneous.

To find an approximate solution, one chooses a finite number of basis (ansatz) functions $b_i(\vec{r})$ to expand the unknown electrostatic potential $\Phi(\vec{r})$ and the charge density $\varrho(\vec{r})$
\begin{align}
\Phi(\vec r) = \textstyle\sum_k \Phi_k b_k(\vec r), \hspace{0.5cm} \varrho(\vec r) = \textstyle\sum_k \varrho_k b_k(\vec r) .
\end{align}

Using the Galerkin approach, the test function $\varphi(\vec{r})$ is approximated using the same basis functions $b_i(\vec{r})$ as for the unknown fields. For a fixed $\varphi(\vec{r})$ the integral on the left-hand side is linear in $\Phi$, allowing us to consider each basis function individually. The relation~\cref{eq:weak_form} is then fulfilled for any $\varphi(\vec{r}) = \sum_i \varphi_i b_i(\vec{r})$.

\begin{align}
 \resizebox{0.88\linewidth}{!}{$\underbrace{\sum_k \Phi_k \int_\Omega \nabla b_i(\vec r) \nabla b_k(\vec r) \, dV}_{\textstyle\sum_k \overline K_{ik} \Phi_k} = \underbrace{\sum_k \varrho_k / \varepsilon \int_\Omega b_i(\vec r) b_k(\vec r) \, dV}_{f_i(\varrho)} $}
\end{align}
Decomposing the domain into a mesh of small sub-domains of, for example, triangular shape and choosing basis functions $b_i(\vec{r})$ that are only non-zero on one of these mesh elements ensures that the matrix $\tens K$ is sparse, which allows for efficient computation.

Choosing polynomials as the ansatz functions on each of these sub-domains allows the exact evaluation of the integrals numerically. In the case of equations with smooth solutions, such as the electrokinetic equations, the accuracy of the FEM approximation benefits more from increasing the polynomial order of the ansatz functions than from increasing the mesh resolution, for the same computational cost.~\cite{guo_h-p_1986}

Ultimately, one is left with a system of linear equations for the coefficients of the solution $\Phi_k$ 
\begin{equation}
\label{eq:fem_discretised}
\tens K \vec \Phi = \vec f(\vec \varrho)
\end{equation}
where $\vec \Phi$ and $\vec \varrho$ are the vectors consisting of all the coefficients $\Phi_k$ and $\varrho_k$, respectively. Applying this procedure to the diffusion-advection and hydrodynamic equations in~\cref{eq:model_summary} yields similar equation systems. The coupling between the different equations is reflected by the fact that both the operator $\tens K$ and the right-hand side $\vec{f}$ can depend on the solutions of the other equations. Combining the resulting discretised equations into one system yields
\begin{equation}
\label{eq:fem_fully_coupled}
\underbrace{\begin{bmatrix}
\tens K_1(\vec \Phi, \vec u) & 0 & 0 \\
0 & \tens K_2 & 0 \\
0 & 0 & \tens K_3
\end{bmatrix}}_{\vec{F}(\vec \Phi, \vec u)}
\begin{bmatrix}
\vec c \\
\vec \Phi \\
\vec u
\end{bmatrix}
=
\begin{bmatrix}
0 \\
\vec f_2(\vec c) \\
\vec f_3(\vec \Phi, \vec c) \\
\end{bmatrix} ,
\end{equation}
where $\tens K_1$ represents the discretised version of all the diffusion-advection equations~\cref{eq:model_concentration}, $\tens K_2$ and $\vec{f}_2$ represent Poisson's equation as previously derived in~\cref{eq:fem_discretised}, and $\tens K_3$ and $\vec{f}_3$ represent Stokes' equations~\cref{eq:model_stokes}. The discrete vector $\vec{c}$ contains all coefficients of the FEM approximation for the ionic concentrations, $\vec{\Phi}$ contains these coefficients for the electrostatic potential, and $\vec{u}$ contains them for the fluid velocity (and pressure).

The discretised and fully coupled equation system~\cref{eq:fem_fully_coupled} is non-linear due to the coupling between the different equations. This equation system can be solved using Newton's method. Obtaining the necessary Jacobian matrix of the differential operator is simple: for a single iteration of Newton's scheme, the fields $\vec \Phi$, $\vec u$, and $\vec c$ are fixed, which makes the operator $\vec{F}(\vec \Phi, \vec u)$ linear and the right-hand side constant, yielding the following iteration scheme
\begin{equation}
\label{eq:fem_newton}
\resizebox{0.88\linewidth}{!}{$\begin{bmatrix}
\tens K_1(\vec \Phi_n, \vec u_n) & 0 & 0 \\
0 & \tens K_2 & 0 \\
0 & 0 & \tens K_3
\end{bmatrix}
\begin{bmatrix}
\vec c_{n+1} - \vec c_n\\
\vec \Phi_{n+1} - \vec \Phi_n\\
\vec u_{n+1} - \vec u_n
\end{bmatrix}
=
\begin{bmatrix}
0 \\
\vec f_2(\vec c_n) \\
\vec f_3(\vec \Phi_n, \vec c_n) \\
\end{bmatrix} $}
\end{equation}

To summarise, the above described procedure and iteration scheme now allow us to guess an initial solution of the electrokinetic equations; in most cases the iteration scheme will then converge to the correct solution. In simulations involving very high surface charges, or a strong imposed voltage bias at the boundaries, the iteration scheme~\eqref{eq:fem_newton} can diverge. An initial guess closer to the actual solution fixes this problem. That is why, in these situations, one slowly ramps up the surface charge or the voltage bias over a series of simulations, using each one's solution as the initial guess for the next.

The flexibility, accuracy, and computational efficiency of the finite-element method have made it an extremely popular choice for the numerical modelling of the electrokinetic equations, amongst others. There are now many powerful and user-friendly simulation codes based on the finite-element method, for example the COMSOL simulation package.

However, in \Cref{sec:results} we show that care must be taken when carrying out finite-element simulations of electrokinetic phenomena due to the occurrence of spurious fluxes and flows, the magnitude of which can easily exceed physical fluxes and flows if not properly controlled. 

In the following section we discuss the origin of spurious fluxes and flows in numerical simulations of electrokinetic phenomena. 

\subsection{The Origin of Spurious Fluxes and Spurious Flows}
\label{sec:origin_spurious}

Unfortunately, discretising the electrokinetic equations~\cref{eq:model_summary} with the standard form of the fluid coupling force~\cref{eq:model_forcedens_estatics} introduces numerical instabilities in the form of so-called spurious fluxes and spurious flows. These artificial currents and flows, if not properly controlled, can exceed physical fluxes and flows, invalidating simulation results. Spurious fluxes appear both in the time-dependent electrokinetic model~\cref{eq:model_summary} and the stationary electrokinetic model (setting $\partial_t = 0$), though the magnitude and shape of these artefacts depend on the specifics of the solver.\\\indent
Spurious flux and spurious flow are caused by the near-cancellation of different contributions to the physical ionic fluxes and physical fluid flows in non-equilibrium situations, and from exact cancellation in equilibrium situations. Due to discretisation errors and limited arithmetic precision, such scenarios involving near/exact-cancellation are hard to treat correctly in numerical schemes.\\\indent
Consider a system in thermodynamic equilibrium where all fluxes and flows vanish. Assuming that the flow velocity actually is zero in the FEM solution, the flux expression~\cref{eq:model_fluxes} and Poisson's equation~\Cref{eq:model_simple_poisson} yield expressions for the polynomial degree of the ansatz functions for the ionic concentrations $[c]$ and the electrostatic potential $[\Phi]$. These expressions read:
\begin{align}
[c]-1 &= [c] + [\Phi]-1 , \\
[\Phi]-2 &= [c] .
\end{align}
\indent The only way to fulfil these equations simultaneously is to set $[c] = -2$ and $[\Phi] = 0$, neither of which is a viable choice: the former because having a polynomial of order $-2$ is impossible; and the latter because a piecewise polynomial of degree $0$ (i.e., a constant) is not compatible with the weak form of the electrokinetic equations typically used in finite-element simulations, which contain a gradient term in the electrostatic potential.
This means that the conditions necessary to eliminate spurious fluxes exactly cannot be realised in an FEM simulation of the type described in~\Cref{sec:fem_solver}. 
Similar issues prevent the pressure gradient and external force~\cref{eq:model_forcedens_estatics} in Stokes' equations~\Cref{eq:model_stokes} from cancelling, which leads to spurious flow. \\\indent 
One way to minimize these effects is to simply increase the grid resolution. This works because polynomials of different order can better approximate one another on smaller domains. The required increase in grid resolution to sufficiently suppress spurious flow and fluxes far exceeds the resolution necessary to accurately treat the gradients in the double layer, and therefore comes with a hefty increase in calculation time, which is extremely undesirable. \\\indent
Despite the fact that in the Capuani \textit{et al.}\cite{capuani_discrete_2004}\ scheme there are no ansatz functions, spurious fluxes still occur due to the near cancellation of the two terms in the diffusive flux expression~\eqref{eq:model_fluxes}. Spurious flows still occur because of the near cancellation between the pressure gradient and the applied force in Stokes' equations~\eqref{eq:model_stokes}.
The coupling of the ionic flux~\eqref{eq:model_fluxes} and the hydrodynamic flow~\eqref{eq:model_stokes} further enhance both these artefacts.

\section{Improvements to the time-dependent FVM, FDM, and LBM solver}
\label{sec:nonlinear_stencil_problem}

Before we move on to the main results of this paper, namely the reduction of spurious flow in simulations of electrokinetic phenomena in both time-dependent solvers and finite-element method solvers, we must first present some results regarding the time-dependent solver based on FVM, FDM, and LBM introduced by~Capuani \textit{et al.}\cite{capuani_discrete_2004}\  and described in Section~\ref{sec:ek_solver}. We will use these improvements when carrying out simulations illustrating issues with spurious flow with the time-dependent solver.

Capuani \textit{et al.}\cite{capuani_discrete_2004}\  proposed a reformulation~\cref{eq:flux_transformed} of the diffusive contribution to the flux expression~\cref{eq:model_fluxes} to reduce spurious fluxes, however, we show here that this reformulation~\cref{eq:flux_transformed} is only strictly equivalent to the diffusive contribution of the physical flux~\cref{eq:model_fluxes} in the continuum limit and leads to higher-order errors in the discretised version~\cref{eq:solver_flux}. These higher-order contributions are in fact unbounded and increase exponentially with the difference in potential between neighbouring cells. This reformulation is therefore undesirable in simulations where large electric gradients appear, e.g., in the simulation of nanopores.

To illustrate this, consider the flux of a single homogeneously distributed species of concentration $c$:
\begin{align}
\label{eq:flux_illustration}
j^\text{diff}_{i}(\vec r) = \frac{D c}{2 \vert \vec{d}_i \vert} & \left( e^{-ze \Phi(\vec{r}) / \kT} + e^{-ze \Phi(\vec{r} + \vec{d}_{i}) / \kT} \right) \times \notag \\
& \left( e^{ze \Phi(\vec{r}) / \kT} - e^{ze \Phi(\vec{r} + \vec{d}_{i}) / \kT} \right) .
\end{align}
By setting $\Delta_{i} \Phi(\vec{r}) = \Phi(\vec{r} + \vec{d}_{i}) - \Phi(\vec{r})$, this can be expressed as:
\begin{align}
\label{eq:flux_sinh}
j^\text{diff}_i(\vec{r}) &= \frac{D c}{\vert \vec{d}_{i} \vert} \sinh \left( -\frac{ze \Delta_{i} \Phi(\vec{r})}{\kT} \right) \\
&= -\frac{D}{\kT} c ze \frac{\Delta_{i} \Phi(\vec{r})}{\vert \vec{d}_{i} \vert} + \mathcal{O} \big( \Delta_{i} \Phi(\vec{r})^2 \big),
\end{align}
from which one recovers the correct expression $\vec{j}^\text{diff} =  - \mu c ze \nabla \Phi$ to first order only. The error grows exponentially with the potential difference of neighbouring cells $\Delta_{i} \Phi$, which results in an additional upper limit for the grid spacing, depending on the maximum of the electric field in the system. \\\indent 
It is worth noting that this higher-order effect only occurs in non-equilibrium situations and does not change the equilibrium distribution, as can be seen by setting the flux in the discrete expression~\cref{eq:solver_flux} to zero and separating the expressions involving the concentrations and the potentials. \\\indent
In order to eliminate this unbounded, exponentially increasing error in the scheme by Capuani \textit{et al.}\cite{capuani_discrete_2004}, we instead employ a direct discretisation of the diffusive part of the physical flux~\cref{eq:model_fluxes} using a finite-difference scheme for the gradient across the links, connecting the nodes at position $\vec{r}$ and $\vec{r} + \vec{d}_{i}$
\begin{multline}
\label{eq:linear_flux}
j^\text{diff}_{ki}(\vec{r}) = -D_{k} \frac{c_{k}(\vec{r}) - c_{k}(\vec{r} + \vec{d}_{i})}{\left| \vec{d}_{i} \right|} \\
 + \mu_{k} z_{k} e \frac{c_{k}(\vec{r} + \vec{d}_{i}) + c_{k}(\vec{r})}{2} \times \frac{\Phi(\vec{r} + \vec{d}_{i}) - \Phi(\vec{r})}{\left| \vec{d}_{i} \right|} .
\end{multline}

In order to illustrate the differences between the scheme by Capuani \textit{et al.}\cite{capuani_discrete_2004}~\cref{eq:solver_flux} and our modified expression~\cref{eq:linear_flux}, we simulate a system consisting of two parallel no-slip plates containing a homogeneous density of a single, uncharged species. A spatially and temporally constant force density acting on the dissolved species along the channel creates a parabolic flow profile. The fact that the dissolved species is uncharged allows it to remain homogeneously distributed, which in turn causes the Fickian diffusion term in the flux density~\cref{eq:model_fluxes}, $-D\nabla c$, to vanish. This system corresponds exactly to the well-known Poisseuille-flow set up; the homogeneous neutral species merely acts as a means to apply a homogeneous force density to the fluid. \\\indent
The fluid's viscosity and the solute's diffusion coefficient were chosen such that the ratio of advective to diffusive/migrative transport is approximately $1$. In this set-up, the potential $\Phi$ is not given by Poisson's equation but instead just prescribed externally.

\begin{figure}
\begin{center}
\includegraphics[width=\linewidth]{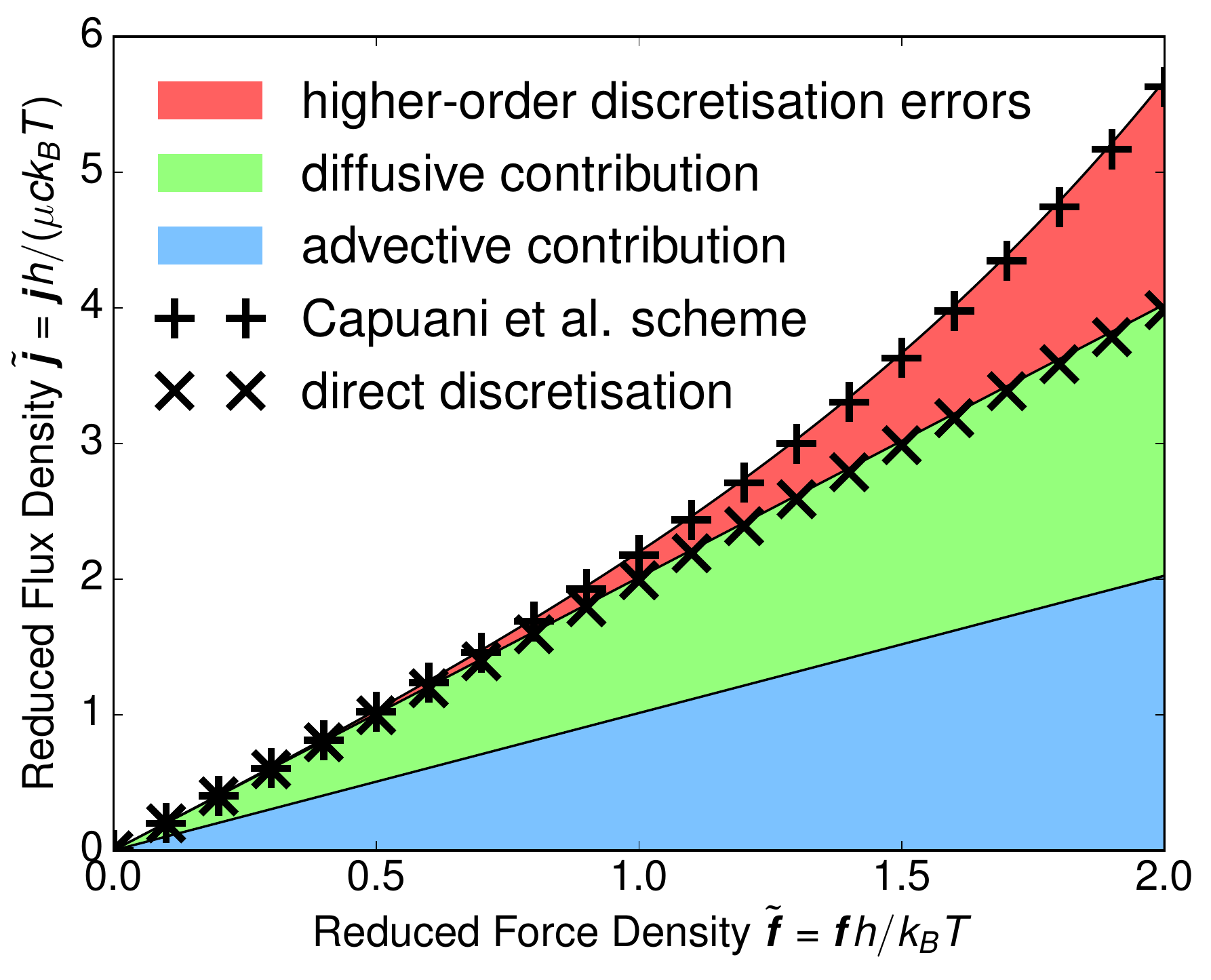}
\end{center}
\caption{Comparison between the Capuani \textit{et al.}\cite{capuani_discrete_2004}\  discretisation~\cref{eq:solver_flux} and our direct discretisation method for the diffusive flux~\cref{eq:linear_flux} in Poiseuille-flow. Theoretical results are shown as coloured areas and split up into contributions coming from advection (blue) and diffusion (green). The higher-order discretisation errors~\cref{eq:flux_sinh} in the Capuani \textit{et al.}\cite{capuani_discrete_2004}\  scheme are shown in red. The crosses ($\times$) show results of simulations with our direct discretisation, the plus symbols (+) show results obtained from simulations using the Capuani \textit{et al.}\cite{capuani_discrete_2004}\  discretisation scheme.}
\label{fig:stencil_comparison}
\end{figure}

\Cref{fig:stencil_comparison} shows how the reduced flux density $\tilde{\vec{j}} = \vec{j} h / (\mu c \kT)$ varies with the reduced force density $\tilde{\vec{f}} = \vec{f} h / \kT$ for the above described Poiseuille-flow system, where $h$ denotes the grid spacing and $\mu$ the solute's mobility. The coloured regions of~\Cref{fig:stencil_comparison} show the analytical results. \\\indent
The blue region shows the advective contribution to the reduced flux, $\tilde{j}^\text{adv}$, which is identical in both our direct discretisation scheme and in the Capuani \textit{et al.}\cite{capuani_discrete_2004}\  scheme, since the fluid motion is solved by a lattice-Boltzmann solver in both cases and the fluid's driving force is unchanged. \\\indent
The green region shows the diffusive contribution, $\tilde{j}^\text{diff}$, to the reduced flux. Our direct discretisation method ($\times$~symbols) reproduces the analytical result for the reduced force densities that we investigated. The diffusive contribution according to Capuani \textit{et al.}\cite{capuani_discrete_2004}\  (+~symbols) shows good agreement with the analytical solution for reduced force densities up to $\tilde{\vec{f}} \approx 0.5$, but begins to deviate for reduced force densities $\tilde{\vec{f}} > 0.5$. In fact, as we showed in~\cref{eq:flux_sinh}, the error between the analytical solution and the numerical solution using the scheme by Capuani \textit{et al.}\cite{capuani_discrete_2004}\  increases exponentially with increasing reduced force density, shown by the red coloured region. \\\indent
It is important to note that in many systems, including the ones investigated by~Capuani \textit{et al.}\cite{capuani_discrete_2004}, the difference is negligible. However, these errors become significant and important in systems involving strong local electric fields, such as the much researched biological and solid state nanopores. These pores typically connect reservoirs of an electrolyte solution otherwise separated by a thin membrane. Due to the high conductivity of the electrolyte, most of the voltage drops at the nanopore, creating very strong local electric fields. Typical measurements include the ionic current through the nanopore as well as the electro-osmotic flow. Both of these are non-equilibrium effects and therefore suffer from the discussed discretisation errors.
In non-equilibrium situations, we therefore recommend that our direct discretisation method~\cref{eq:linear_flux} be used. \\\indent 
In \Cref{sec:results_sims}, we carry out several simulations using the time-dependent solver based on FVM, FDM, and LBM introduced by~Capuani \textit{et al.}\cite{capuani_discrete_2004}\  and described in Section~\ref{sec:ek_solver}. When carrying out these simulations, we use the direct discretisation method~\cref{eq:linear_flux} that we propose here, rather than the flux reformulation method~\cref{eq:solver_flux} presented by~Capuani \textit{et al.}\cite{capuani_discrete_2004}.

\section{Simulation Setup and Methods}
\label{sec:methods}

To illustrate the issues with spurious fluxes and flows, we simulate two systems. The first is a commonly encountered system, a charged sphere at rest immersed in an electrolyte solution.
 
Section~\ref{sec:time_dependent_methods} describes the simulation set up and parameters we use when carrying out time-dependent simulations based on FVM, FDM, and LBM introduced by~Capuani \textit{et al.}\cite{capuani_discrete_2004}\ and described in Section~\ref{sec:ek_solver}, incorporating our direct discretisation improvements as described in Section~\ref{sec:nonlinear_stencil_problem}. Section~\ref{sec:fem_methods} describes the simulation setup and parameters used in finite-element simulations of the charged sphere system.

The second system we simulate is a charged nanopore. Section~\ref{sec:nanopore_methods} describes the simulation setup and methods of a charged nanopore system, which we simulate to illustrate how spurious fluxes and flows can invalidate simulation results. For this system, we only carry out finite-element simulations. 

\subsection{Time-Dependent Simulations}
\label{sec:time_dependent_methods}

To solve the system comprising a charged sphere at rest immersed in an electrolyte solution numerically, we deviate from the approach of Capuani \textit{et al.}\cite{capuani_discrete_2004}\ by using a method based on discrete Fourier transforms (DFT) to solve the discretised Poisson equation~\cref{eq:model_stencil} instead of a successive-over-relaxation scheme. This solver delivers superior precision at comparable computational cost for moderate system sizes and scales more favourably ($\mathcal{O}(n \log n)$ instead of $\mathcal{O}(n^{4/3})$). The one drawback of this method compared with the successive-over-relaxation scheme is that the electric permittivity must be homogeneous. 

We use fast Fourier transforms (FFT) to obtain the DFT $\hat{\varrho}(\vec{r})$ of the charge distribution, then multiply with the DFT of the exact Green's function for the discrete Poisson problem~\cref{eq:model_stencil} to obtain the DFT of the electrostatic potential $\hat{\Phi}(\vec{r})$, which assumes the following form:
\begin{align}
\label{eq:methods_electrostatics_phik}
\hat{\Phi}(\vec{k}) & = -\frac{2 \pi  h^{2} \lB \kT}{\displaystyle \left[ \sum_{j\in\{x,y,z\}} \cos \left( \frac{2 \pi k_{j}}{N_{j}} \right) \right] - 3} \hat{\varrho}(\vec{k}), \vec{k} \neq 0 \\
\label{eq:methods_electrostatics_phi0} \hat \Phi(0) & = 0 .
\end{align}
We obtain the desired real space representation of the electrostatic Potential $\Phi(\vec{r})$ using an inverse FFT.
While this method's superior numerical precision helps maintain momentum conservation in situations involving moving boundaries, it does not significantly influence spurious fluxes and spurious flow in this system containing a fixed boundary \ie all of the problems of spurious fluxes and flows remain. 

We employ a two-relaxation-time LB scheme as opposed to the single-relaxation-time LB scheme used by Capuani \textit{et al.}\cite{capuani_discrete_2004}. In the low Mach number and low Reynolds number limit realized in this investigation, both LB methods reproduce the incompressible Stokes equations~\eqref{eq:model_stokes}.\cite{fischer_raspberry_2015, graaf_raspberry_2015}

The specifics of the investigated system enter the simulation as boundary conditions. The fluid velocity, as well as the normal flux of ionic species at the sphere's surface is zero (no-slip and impermeable). We introduce the sphere's surface charge by charging the outermost layer of the boundary nodes accordingly. The simulation domain is a cubic box, whose outer boundaries are periodic in all directions. Our own numerical experiments have shown that a distance of $5\lD$, between the sphere's surface and the periodic boundary is sufficient to eliminate finite-size effects, if the system is charge neutral. We ensure charge neutrality by adding the necessary amount of counter-ions in addition to the salt ions. $\lD = 1/\sqrt{4\pi\lB \sum_k z_k^2 \bar c_k}$ denotes the Debye length. We use two oppositely charged monovalent ionic species at equal concentrations of $\bar c_{1,2} = 1\U{mmol/l}$ resulting in a Debye length of $\lD = 9.7\U{nm}$ for a Bjerrum length of $\lB = 0.7\U{nm}$ (water at $300\U{K}$), and a sphere of radius $10\U{nm}$ with a surface charge $\sigma = -0.03\U{C/m^2}$. These are typical parameters for experimental systems. The grid resolution is $h = 1\U{nm}$, to resolve the Debye length.

\subsection{Finite-Element Simulations}
\label{sec:fem_methods}

The parameters and boundary conditions in the finite-element simulation of the stationary electrokinetic equations~\cref{eq:model_summary} (setting $\partial_t = 0$) are the same as described above. The only difference we make is at the outer boundary, where we model a bulk fluid instead: we set the ionic species' concentrations to their bulk value, the electrostatic potential to zero, and require the normal stress of the fluid to vanish. Setting the normal stress to zero implies no momentum exchange with the outer boundary, but still allows fluid flow through the boundary. This is possible because momentum is only transported through viscous friction and not through convection in Stokes' equations~\cref{eq:model_stokes}. Charge neutrality is not enforced explicitly but is achieved through the coupling with the reservoir at the outer boundary for sufficiently large simulation domains.

We take advantage of the fact that the domain can be divided non-uniformly and use much smaller mesh elements in the region close to the sphere's charged surface, where we expect strong gradients in all of the fields.
At the sphere surface, we place mesh elements of size $\lD/20$ and gradually increase the element size to $\lD/4$ at a distance of $90\U{nm}$ from the sphere's surface at the outer domain boundary, where we expect bulk-like behaviour. We can afford such a high resolution by taking advantage of the cylindrical symmetry of the system. As ansatz functions, we use piecewise polynomials of degree 3 for the ionic concentrations, degree 2 for the electrostatic potential, and degree 2 and 1 for the fluid velocity and pressure, respectively.

\subsection{Simulations of a Nanopore}
\label{sec:nanopore_methods}

\begin{figure}
\includegraphics[width=\linewidth]{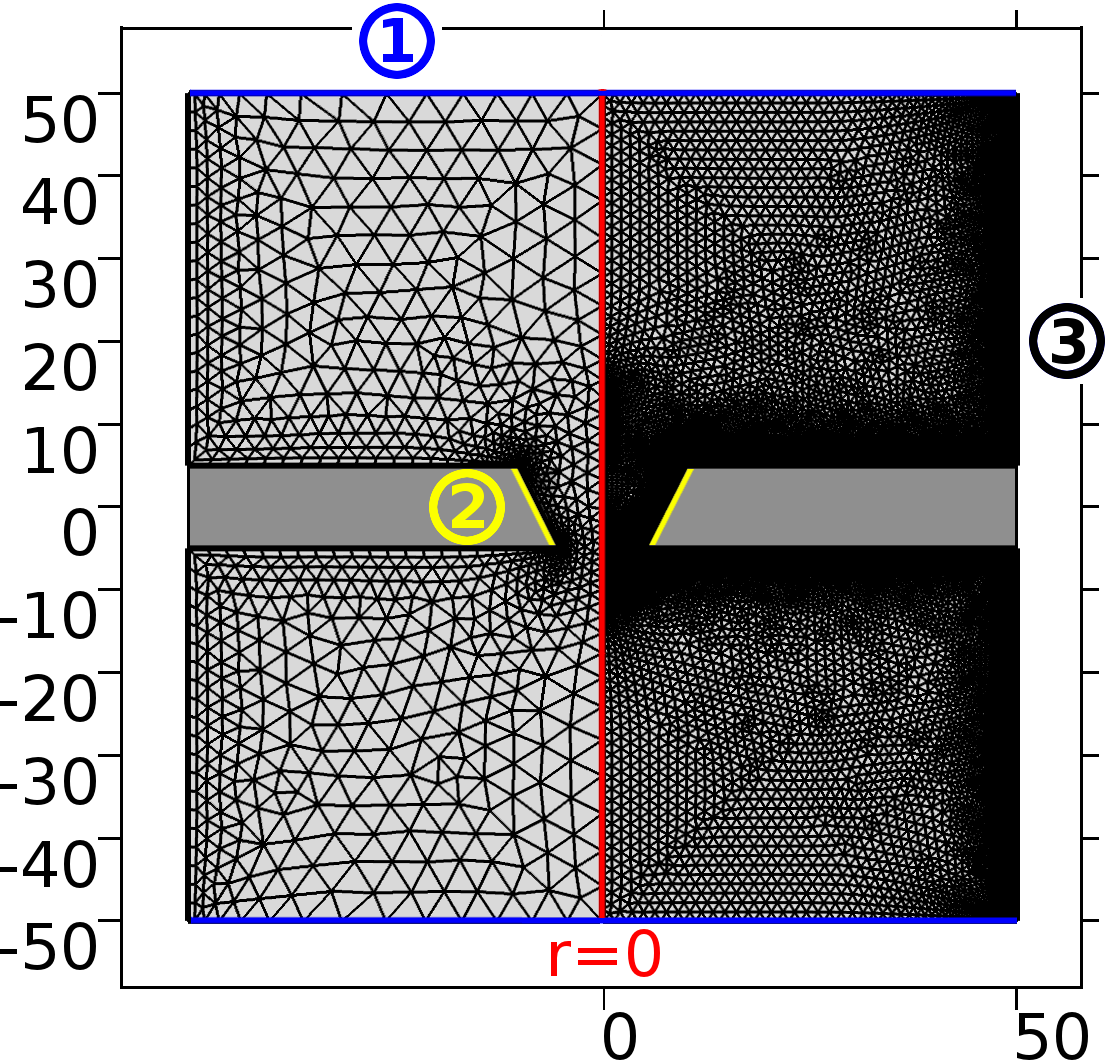}
\caption{The geometry, boundary conditions, and mesh used for the charged nanopore system. The bulk boundary conditions (No.\ 1. blue) are applied at the top and bottom of the simulation domain. The electro-osmotic flow is created at the inner boundaries of the nanopore (No.\ 2, yellow), which are the only charged boundaries in the system. The remaining boundaries (No.\ 3, black) are uncharged, no-slip boundaries impermeable to ions. Red denotes the symmetry axis. The left half of the image depicts the coarse grid used, resulting in an equation system with $26,895$ degrees of freedom, while the right half depicts the fine mesh used resulting in an equation system with $367,431$ degrees of freedom.}
\label{fig:mesh_nanopore}
\end{figure}

We simulate a charged nanopore system that comprises two electrolyte reservoirs connected by a pore in the shape as depicted in \Cref{fig:mesh_nanopore}. The inner boundaries of the nanopore (No.\ 2, yellow) are negatively charged with a surface charge $\sigma = -0.01\U{C/m^2}$. We apply a potential difference to the pore by setting the electrostatic potential to $0\U{V}$ at the lower bulk boundary condition (No.\ 1, blue) and to a non-zero voltage at the upper bulk boundary. We use the same condition of no normal stress for the fluid as in the sphere system described in \Cref{sec:fem_methods}. The remaining boundaries at the membrane separating the reservoirs and along the sides of the simulation domain (No.\ 3, black) are uncharged (vanishing normal electric field), no-slip boundaries impermeable for ions (normal flux vanishing). As before, we take advantage of the rotational symmetry of the system.
We use the same ansatz functions and remaining parameters as for the system previously described in \Cref{sec:fem_methods} and prepare two different meshes. The first is a coarse mesh, with significant refinement only at the corners of the charged walls, which otherwise contains mesh elements of sizes between $\lD/5$ and $\lD/2$. The second is an extremely fine mesh covering the whole simulation domain with elements of size $\lD/10$ and even smaller mesh elements at all but the bulk boundaries. These two meshes result in discretised equation systems with $26, 895$ and $367, 431$ degrees of freedom, respectively.

In \Cref{sec:results_sims_nanopore}, we will show that one obtains similar results for the coarse and fine grid simulation if one implements the improved fluid coupling~\cref{eq:model_forcedens_friction}, which we introduce in \Cref{sec:corrected_coupling}, and that the results differ wildly using the traditional fluid coupling~\eqref{eq:model_forcedens_estatics}.

\section{Results}
\label{sec:results}

In order to reduce the magnitude of spurious flow, we propose a correction to the fluid coupling force~\cref{eq:model_forcedens_estatics} in~\Cref{sec:corrected_coupling}, which does not eliminate spurious flow entirely, but decreases it by several orders of magnitude without increasing the computational cost, in contrast with the transformation~\cref{eq:solver_flux} involving exponentials employed by~Capuani \textit{et al.}\cite{capuani_discrete_2004} We demonstrate that in typical applications, this reduction of the magnitude of the artefacts is sufficient to obtain correct results in simulations of electro-osmotic flow and electrophoresis.

\subsection{Theory}
\label{sec:corrected_coupling}

To address this issue of spurious flow, instead of the flux term~\cref{eq:model_fluxes}, we modify the hydrodynamic driving force~\cref{eq:model_forcedens_estatics} so as to minimize the spurious flow contribution to the fluid velocity field $\vec{u}$. Note that we have significant freedom in choosing this force term $\vec{f}$ in Stokes' equations~\cref{eq:model_stokes}, as any force term modified only by a gradient field will not change the resulting velocity field $\vec{u}$. To explain this, consider Stokes' equations:
\begin{equation}
\begin{aligned}
\eta \bm{\nabla^{2}} \vec{u} &= \nabla p - \vec{f}, \\
\nabla \cdot \vec{u} &={} 0 .
\end{aligned}
\tag{\ref{eq:model_stokes}}
\end{equation}
Note that both the velocity $\vec{u}$ and the pressure $p$ are unknowns and that $p$ needs to be chosen in such a way, that the solution for the velocity field $\vec{u}$ also fulfils the incompressibility constraint. For appropriately chosen boundary conditions, the solutions for $\vec{u}$ and $p$ are unique (apart from a constant offset in the pressure).
We can therefore add any gradient field $\nabla \omega$ to the force density
\begin{align}
\label{eq:model_stokes_grandientfield}
\eta \bm{\nabla^{2}} \vec{u} &= \nabla (\underbrace{p' - \omega}_{p}) - \vec{f} ,
\end{align}
because this gradient field can be absorbed into the pressure gradient. Due to the uniqueness of solutions, this modified pressure field $p'-\omega$ has to match the solution for the pressure $p$ from the unmodified Stokes' equations~\cref{eq:model_stokes}, and the velocity field $\vec{u}$ remains unchanged.

To reduce the issues with spurious flow, we propose an extension of the hydrodynamic driving force~\cref{eq:model_forcedens_estatics} in the form of a gradient field corresponding to the ionic species' ideal gas pressure. The hydrodynamic driving force extended in this way reads:
\begin{align}
\label{eq:model_forcedens_friction}
\vec{f} = -\textstyle\sum_{k} ( \kT \nabla c_{k} + z_{k} e c_{k} \nabla \Phi ) .
\end{align}
The advantage of this choice is that the force density acting on the fluid vanishes in equilibrium. To explain this property, note that by definition, both the flow velocity and the ionic fluxes vanish in equilibrium
\begin{align}
\label{eq:model_equilibrium_flux}
\vec{j}_{k}/\mu_{k}  &={} -\kT \nabla c_{k} - z_{k} e c_{k} \nabla \Phi = 0 ,\\
\label{eq:model_equilibrium_flow}
\vec{u} &= 0 .
\end{align}
Stokes' equations~\cref{eq:model_stokes} for this equilibrium situation reveal that the modified force density~\cref{eq:model_forcedens_friction} does not lead to a pressure build-up
\begin{equation}
\label{eq:drivingforce_vanish_corrected}
0 = \nabla p + \underbrace{\textstyle\sum_{k} ( \kT \nabla c_{k} + z_{k} e c_{k} \nabla \Phi )}_{=0} .
\end{equation}
The common coupling force~\cref{eq:model_forcedens_estatics} on the other hand, does not vanish in equilibrium and must be countered by a pressure build-up in the fluid to fulfil the condition of zero flow velocity
\begin{equation}
\label{eq:drivingforce_vanish_traditional}
0 = \nabla p + \textstyle\sum_{k} z_{k} e c_{k} \nabla \Phi .
\end{equation}
This effect is especially strong in electric double layers, where the charge density and electrostatic potential gradients are large.  In numerical schemes, exact cancellation, especially when gradients are involved, is usually problematic. If the discretisation does not allow for this to be fulfilled exactly, spurious flow \emph{must} occur. But even if the discretisation scheme would allow for this cancellation, numerical errors can still cause spurious flow.

Finally, we should remark that our proposed corrected coupling force~\cref{eq:model_forcedens_friction} can be expressed via the diffusive flux from~\eqref{eq:model_fluxes} or the chemical potential $\nu_k = \kT \log(\Lambda_k^3 c_k) + z_k e \Phi$, with the thermal de Broglie wavelength $\Lambda_k$
\begin{align}
\label{eq:model_forcedens_interpretation}
\vec{f} = \textstyle\sum_{k} \vec{j}^\text{diff}_{k}/\mu_{k} = -\textstyle\sum_{k} c_k \nabla \nu_k.
\end{align}
The first equality allows one to interpret the modified force density~\cref{eq:model_forcedens_friction} as a friction coupling, since the diffusive flux normalised by the ions' mobility $\vec{j}^\text{diff}_{k}/\mu_{k}$ represents exactly the drag force acting between the ions and the fluid. The second equality demonstrates that our proposed modified force density~\cref{eq:model_forcedens_friction} is the net thermodynamic driving force, which must vanish in equilibrium.

Using the modified force density, the hydrodynamic equations read
\begin{align}
\label{eq:model_stokes_forceV2}
\eta \bm{\nabla^{2}} \vec{u} &= \underbrace{ \nabla p' + \textstyle\sum_{k} ( \kT \nabla c_{k} }_{= \nabla p} + z_{k} e c_{k} \nabla \Phi ) .
\end{align}

\subsection{Numerical Simulations of a Charged Sphere in an Electrolyte}
\label{sec:results_sims}

To demonstrate the improvement offered by our new coupling force, we simulate a charged sphere in an electrolyte solution using both the time-dependent~Capuani \textit{et al.}\cite{capuani_discrete_2004}\ scheme, as described in Section~\ref{sec:ek_solver} and~\ref{sec:time_dependent_methods}, and the FEM solver, as described in Section~\ref{sec:fem_solver} and~\ref{sec:fem_methods}.  

In this system, ions of opposite charge to the sphere should accumulate in a diffuse layer in the vicinity of the sphere's surface. After a period of time, all ions will have rearranged into an equilibrium configuration. In the absence of any external forces, the system should then be completely at rest.

This exact, theoretical solution for both methods requires the fluid pressure gradient to cancel the fluid coupling force, as demonstrated by~\cref{eq:drivingforce_vanish_traditional} for the traditional coupling force~\cref{eq:model_forcedens_estatics}, and by~\cref{eq:drivingforce_vanish_corrected} for the corrected coupling~\cref{eq:model_forcedens_friction}. Due to discretisation errors, spurious fluxes and flows occur, as described in Section~\ref{sec:origin_spurious}. 

\begin{figure}
\begin{center}
\includegraphics[height=0.7\linewidth]{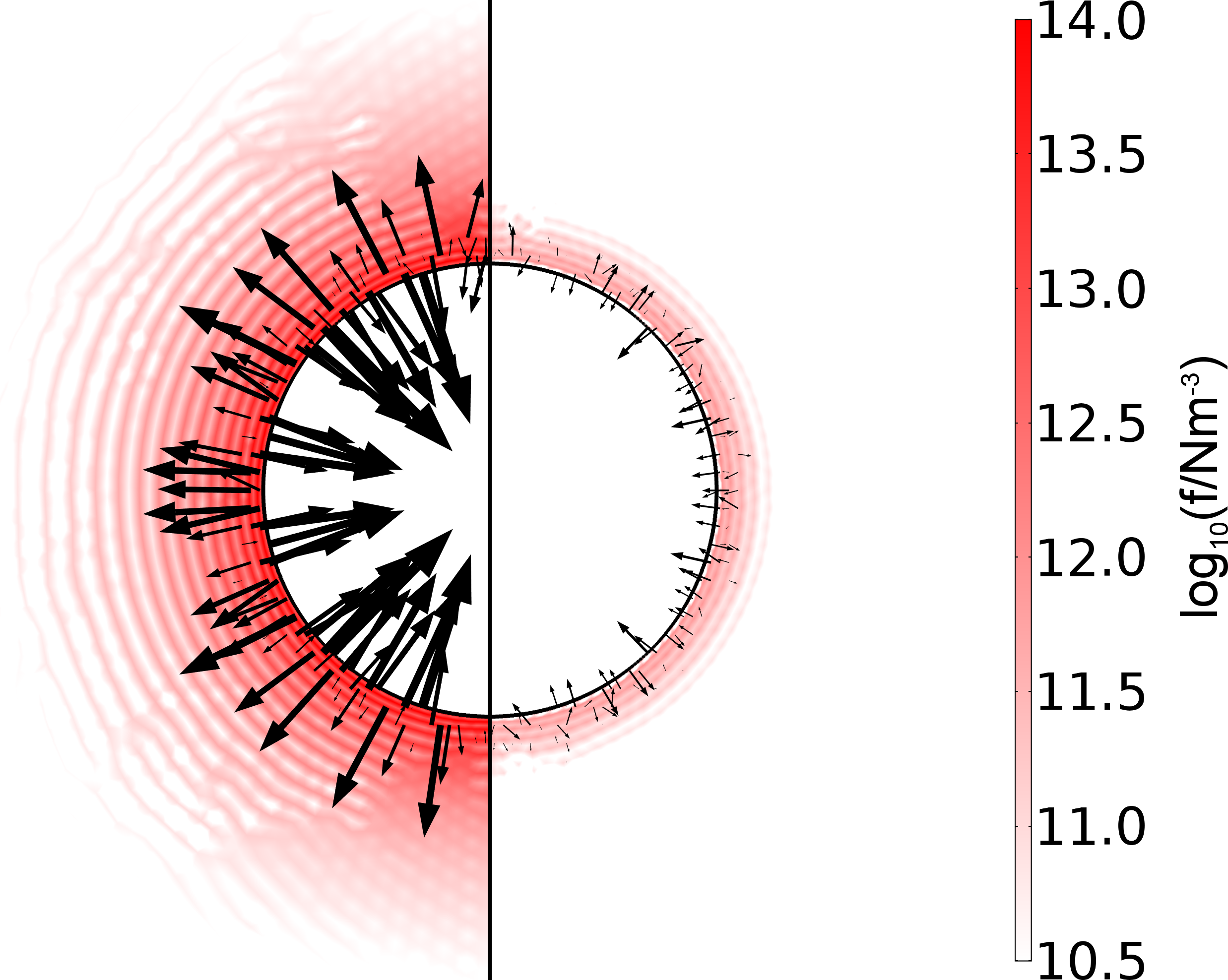}
\end{center}
\caption{{\bf FEM Solver}. The image compares the cancellation error between the fluid coupling force and the pressure gradient (black arrows at randomly selected positions), as well as the magnitude of this error (shades of red) for the simulation using the traditional fluid coupling~\cref{eq:model_forcedens_estatics} (left half) and the improved coupling ~\cref{eq:model_forcedens_friction} (right half). In this equilibrium situation, the gradient of the fluid pressure and the coupling force should cancel each other exactly. The magnitude of the cancellation errors are 115 times smaller using our improved coupling scheme (right half). }
\label{fig:spurious_force_fem}
\end{figure}

Figure~\ref{fig:spurious_force_fem} shows the comparison between the traditional force coupling term~\cref{eq:model_forcedens_estatics} and our improved force coupling term~\cref{eq:model_forcedens_friction} using FEM simulations as described in Section~\ref{sec:fem_solver} and~\ref{sec:fem_methods}. The left half of the figure shows the result for the traditional scheme~\cref{eq:model_forcedens_estatics}, and the right-hand side for our modified scheme~\cref{eq:model_forcedens_friction}.

On the left-hand side, the black arrows show the cancellation error between the gradient of the fluid pressure and the traditional coupling force~\cref{eq:model_forcedens_estatics} in the Stokes' equations~\eqref{eq:model_stokes}, and the red colour shows the magnitude of this cancellation error. The magnitude of this error varies from $2.45 \times 10^{2}\U{Nm}^{-3}$ in bulk to $1.76 \times 10^{14}\U{Nm^{-3}}$ at the charged surface. 

The right-hand side shows our corrected fluid coupling scheme~\cref{eq:model_forcedens_friction}. The magnitude of the cancellation errors are reduced by a factor of $115$ to values between $7.28 \times 10^{2}\U{Nm}^{-3}$ and $1.53 \times 10^{12}\U{Nm^{-3}}$ using our improved coupling scheme.

\begin{figure}
\begin{center}
\includegraphics[height=0.7\linewidth]{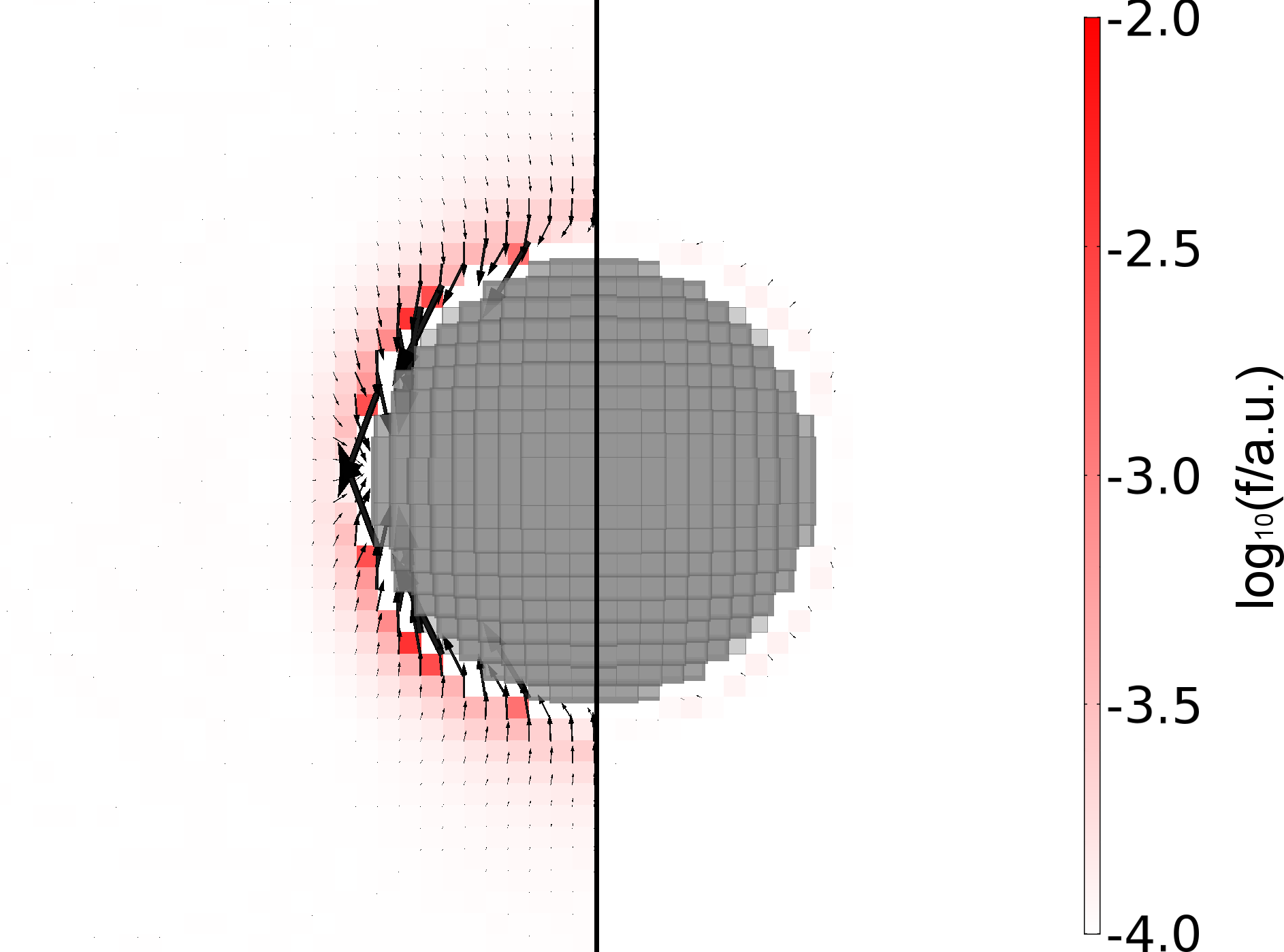}
\end{center}
\caption{{\bf Time-Dependent LB-based Solver}. The image compares the cancellation error between the fluid coupling force and the pressure gradient (black arrows), as well as the magnitude of this error (shades of red) for the simulation using the traditional fluid coupling~\cref{eq:model_forcedens_estatics} (left half) and the improved coupling ~\cref{eq:model_forcedens_friction} (right half). For reasons discussed in \Cref{sec:results_sims}, the cancellation error cannot be evaluated right at the sphere's surface, where it assumes its maximum, which is why this image understates the reduction of the cancellation error. Our improved scheme (right half) shows a factor of $25$ reduction in the cancellation error, but as we discuss in the text, the true reduction factor is $125$.}
\label{fig:spurious_force_ek}
\end{figure}

\Cref{fig:spurious_force_ek} shows the same system using the time-dependent solver described in \Cref{sec:ek_solver} and \ref{sec:time_dependent_methods}. The left half of the figure implements the traditional fluid coupling scheme~\cref{eq:model_forcedens_estatics}, and the right-hand side our improved coupling scheme~\cref{eq:model_forcedens_friction}.

The magnitude of the cancellation errors using the traditional coupling scheme (left half) range from $9.18 \times 10^{-10}\U{a.u.}$ to $7.46 \times 10^{-3}\U{a.u.}$ (a.u.\ = arbitrary units). Using our improved force coupling, the magnitude of the errors vary from $8.69 \times 10^{-9}\U{a.u.}$ to $3.82 \times 10^{-4}\U{a.u.}$, which represents a reduction by a factor of $20$. However, the true reduction in the cancellation error is actually much larger than this because the largest errors occur at the sphere's surface, the value of which we are unable to evaluate due to difficulties in calculating the pressure gradient at a no-slip boundary using the Capuani \textit{et al.}\cite{capuani_discrete_2004}\ scheme. We show later in this section that the true reduction in the cancellation error is a factor of $125$.

Due to the linearity of Stokes' equations~\cref{eq:model_stokes}, the spurious flow decreases by the same factor as the cancellation errors shown in Figure~\ref{fig:spurious_force_fem} and \ref{fig:spurious_force_ek}. Therefore, we expect a two order of magnitude reduction in the spurious flow velocity using our improved scheme~\cref{eq:model_forcedens_friction} as compared with the traditional coupling scheme~\cref{eq:model_forcedens_estatics}. It should be noted that the converse is true: a reduction in spurious flow of a factor $\alpha$ implies a reduction in the cancellation error by the same factor $\alpha$.

\begin{figure}
\begin{center}
\includegraphics[height=0.7\linewidth]{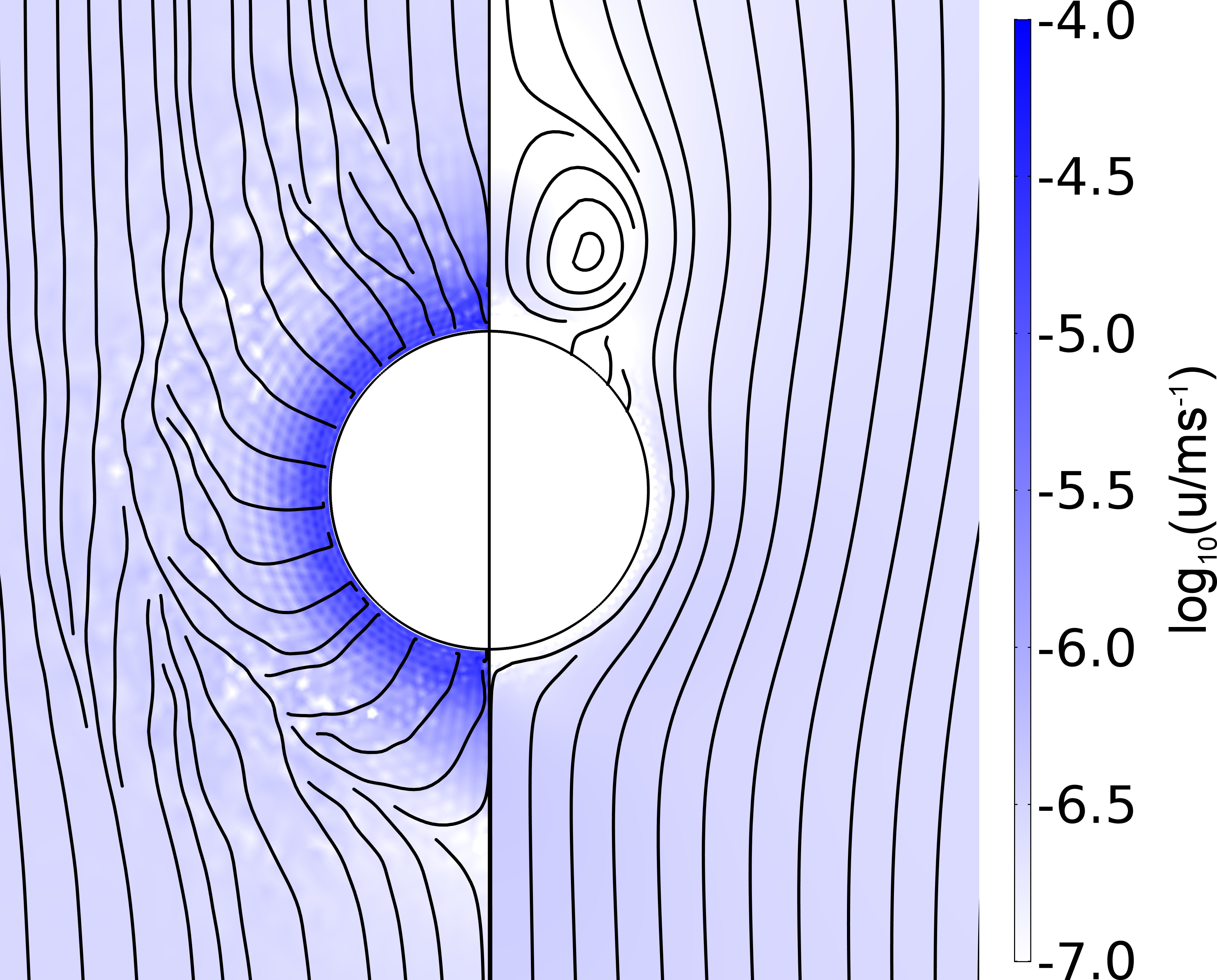}
\end{center}
\caption{{\bf FEM Solver}. Comparison of the spurious fluid flow velocity (black flow lines) corresponding to Fig.~\ref{fig:spurious_force_fem} and its magnitude (blue colour) with the conventional fluid coupling force~\cref{eq:model_forcedens_estatics} (left half) and the corrected coupling force~\cref{eq:model_forcedens_friction} (right half). The artefacts in the flow velocity are reduced by a factor of 115 when using the improved coupling scheme.}
\label{fig:spurious_flow_fem}
\end{figure}

Figure~\ref{fig:spurious_flow_fem} shows the flow fields obtained from the FEM simulation for the traditional coupling~\cref{eq:model_forcedens_estatics} (left-hand side) and our improved coupling~\cref{eq:model_forcedens_friction} (right-hand side). The exact, theoretical solution is a fluid at rest. The spurious flow velocity using the traditional scheme~\cref{eq:model_forcedens_estatics} varies from $0\U{m/s}$ to $7.83 \times 10^{-5}\U{m/s}$ at the sphere's surface, while in our improved scheme~\cref{eq:model_forcedens_friction} (right-hand side), the maximum spurious flow velocity varies from $0\U{m/s}$ to $5.01 \times 10^{-7}\U{m/s}$, which is the 2 orders of magnitude reduction that we expect.

The black flow lines denote the shape of the flow field, while the blue colour denotes the flow magnitude. Since these are simulations of an equilibrium system, the exact solution for the flow field is a fluid at rest. \Cref{fig:spurious_flow_fem} shows that, as expected, the spurious flow velocity for the corrected fluid coupling~\cref{eq:model_forcedens_friction} (right-hand side) is reduced by the same factor as the cancellation error between the pressure gradient and the coupling force. For comparison, the flow resulting from the simulations with the traditional fluid coupling~\cref{eq:model_forcedens_estatics} is shown on the left-hand side: there is a two orders of magnitude difference in the spurious flow velocity between the traditional coupling and our improved coupling.

\begin{figure}
\begin{center}
\includegraphics[height=0.7\linewidth]{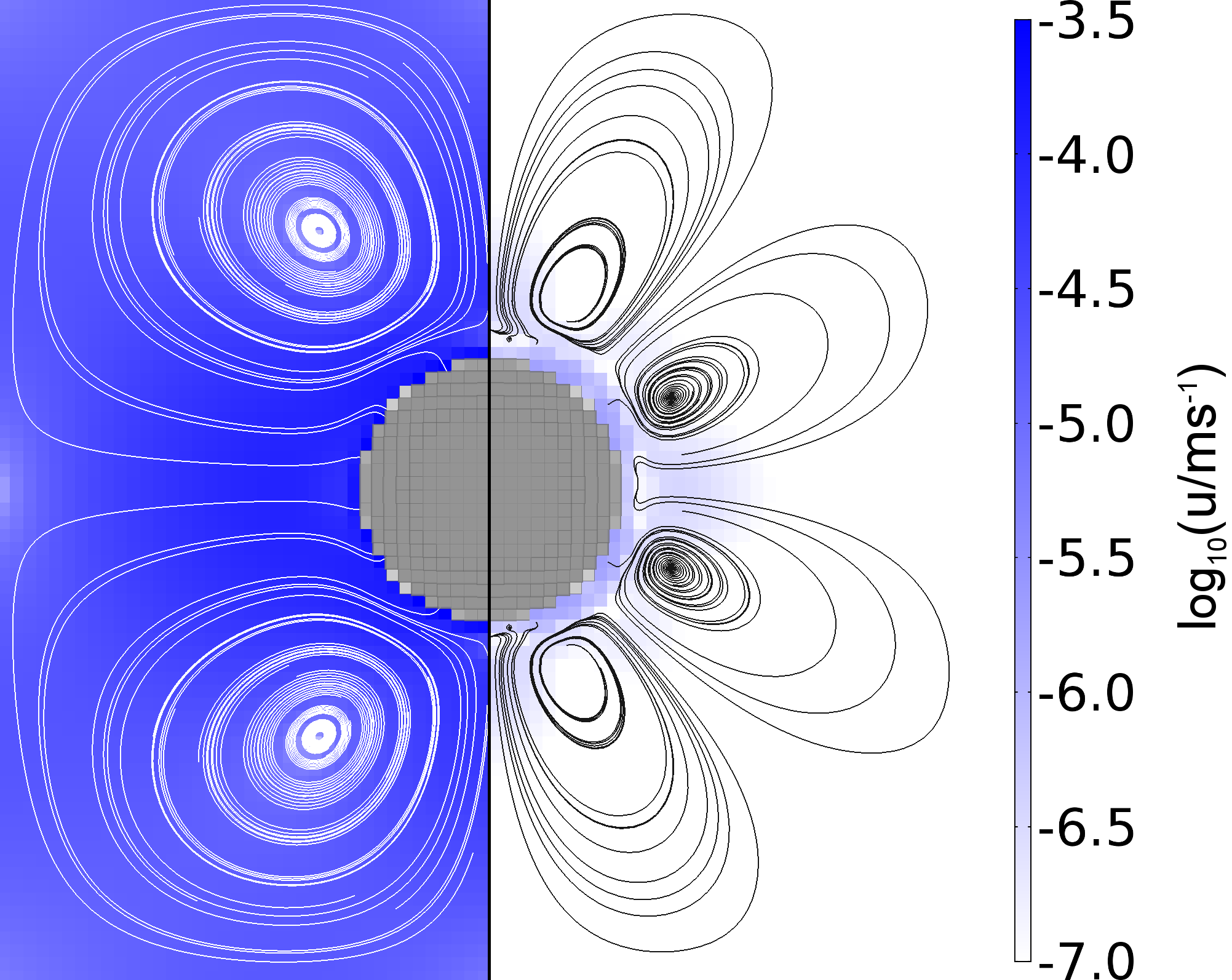}
\end{center}
\caption{{\bf Time-Dependent LB-based Solver}. Comparison of the spurious fluid flow velocity (white/black flow lines) corresponding to Fig.~\ref{fig:spurious_force_ek} and its magnitude (blue colour) with the conventional fluid coupling force~\cref{eq:model_forcedens_estatics} (left half) and the corrected coupling force~\cref{eq:model_forcedens_friction} (right half). The artefacts in the flow velocity are reduced by a factor of 125 when using the improved coupling scheme.}
\label{fig:spurious_flow_ek}
\end{figure}

The flow field from the simulations using the Capuani \textit{et al.}\cite{capuani_discrete_2004}\ scheme and corresponding to Figure~\ref{fig:spurious_force_ek} is shown in Figure~\ref{fig:spurious_flow_ek}. Using the traditional coupling~\cref{eq:model_forcedens_estatics} (left-hand side), the spurious flow velocity varies from $0$ to $2.5 \times 10^{-4}\U{a.u.}$, while our improved scheme~\cref{eq:model_forcedens_friction} shows a two order of magnitude reduction and the maximum spurious flow velocity is of the order of $2.0 \times 10^{-6}\U{a.u.}$

Therefore, using the Stokes' equation linearity argument above, we know that the true force cancellation error is instead a factor of $125$ smaller in Figure~\ref{fig:spurious_force_ek} using our improved force coupling, rather than the factor of $25$ that Figure~\ref{fig:spurious_force_ek} shows. This is because we expect the largest cancellation errors to occur at the sphere's surface (as shown in Figure~\ref{fig:spurious_force_fem}), a region in which we were unable to evaluate the pressure gradient for the Capuani \textit{et al.}\cite{capuani_discrete_2004}\  scheme.

To summarise, our improved coupling scheme provides a factor of $115$ reduction in the magnitude of the force cancellation errors and in the spurious flow velocity using FEM simulations, and a factor of $125$ reduction using the Capuani \textit{et al.}\cite{capuani_discrete_2004}\ time-dependent solver. It should be noted that in the FEM solver our reduction in the cancellation error is achieved with identical computational cost, and in the original Capuani~\textit{et al.}\cite{capuani_discrete_2004}\ with reduced computational cost. 

\subsection{Finite-Element Simulations of a Nanopore}
\label{sec:results_sims_nanopore}

Finally, to highlight how spurious flow can produce wrong simulation results when the spurious flows are of the same magnitude as physical flows, we carry out FEM simulations of the nanopore system introduced in \Cref{sec:nanopore_methods}.

We apply typical bias voltages in the range of $-100\U{mV}$ to $100\U{mV}$ between the upper and lower reservoir boundary (No.\ 1 in \Cref{fig:mesh_nanopore}) and measure the net fluid flow through the nanopore. The flow is caused exclusively by the applied voltage difference through electro-osmosis, as we keep the pressure at the upper and lower reservoir boundary the same. The electro-osmotic flow happens at the inner pore boundaries (No.\ 2 in \Cref{fig:mesh_nanopore}) -- the only charged boundaries in the system. These boundaries are negatively charged, and the electro-osmotic flow is therefore oriented in the direction of the electric field. We expect flow in the positive direction for positive voltage bias, no flow in the case of no bias, and flow in the negative direction for negative voltage bias. This is because there is an excess of positive ions in the double layer in the vicinity of the charged nanopore surface. These positive ions move in the direction of the applied electric field, in turn driving fluid flow in the same direction.

\begin{figure}
\begin{center}
\includegraphics[width=\linewidth]{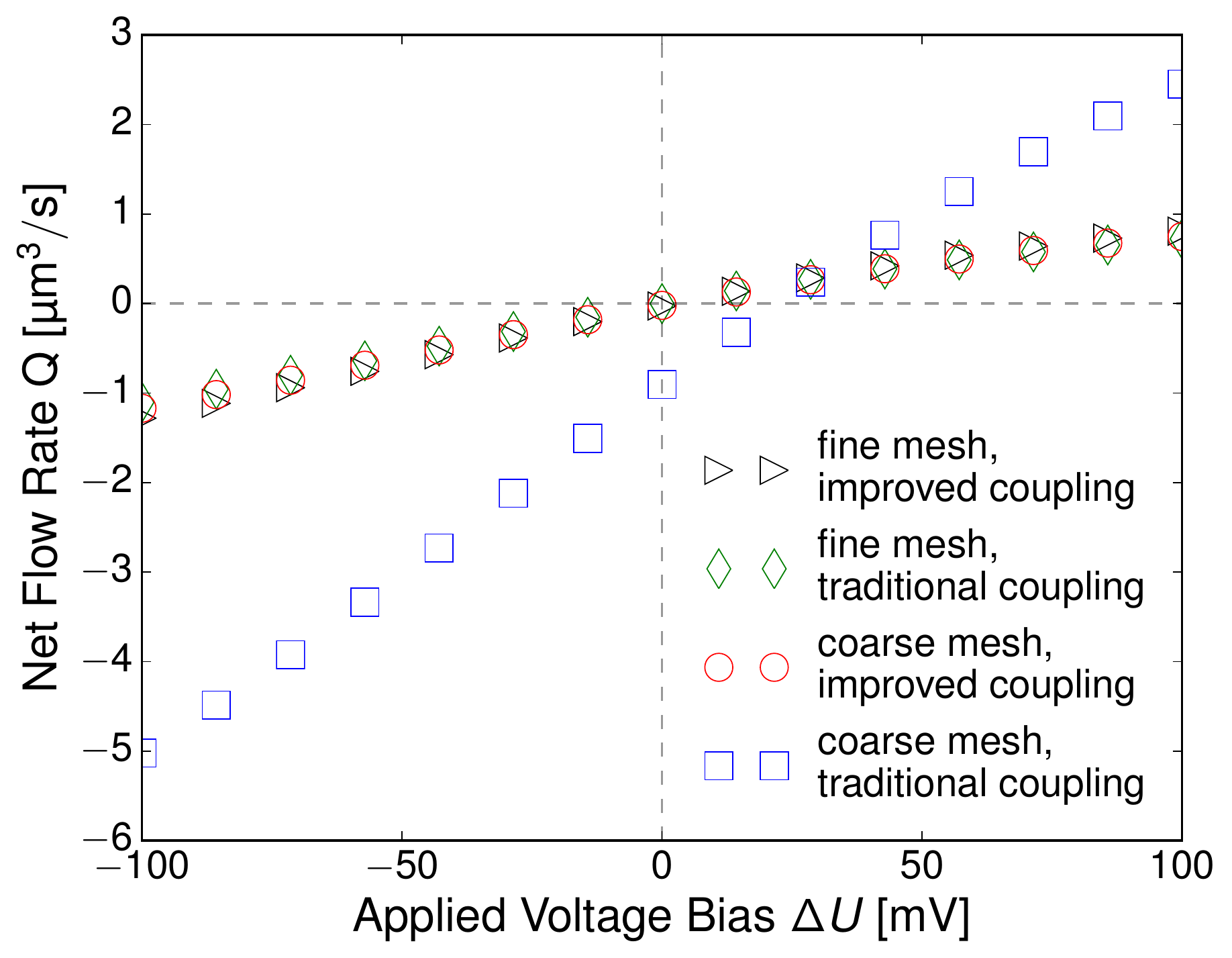}
\end{center}
\caption{Net electro-osmotic fluid flow through the nanopore as depicted in \Cref{fig:mesh_nanopore} for applied bias voltages between $-100\U{mV}$ and $100\U{mV}$. Different symbols depict different combinations of the coarse and fine mesh with the traditional and improved fluid coupling, as explained in the legend. Simulations using the improved fluid coupling~\cref{eq:model_forcedens_friction} and the coarse mesh reproduce the fine-grid solutions, while simulations using the traditional fluid coupling~\cref{eq:model_forcedens_estatics} on a coarse mesh are dominated by spurious flow and yield massively incorrect results.}
\label{fig:nanopore_netflow}
\end{figure}

\Cref{fig:nanopore_netflow} shows the net fluid flow through the pore as a function of the applied voltage bias for all combinations of coarse and fine mesh, as well as traditional~\cref{eq:model_forcedens_estatics} and improved~\cref{eq:model_forcedens_friction} fluid coupling. For the combination of (i) coarse mesh and the traditional fluid coupling~\cref{eq:model_forcedens_estatics} (blue squares), spurious flow dominates the fluid behaviour, leading to completely incorrect results, including flow in equilibrium: for zero applied voltage bias, the net flow rate should be zero, which is not the case. 

A common technique to reduce spurious flow is to increase the grid resolution, at significant computational cost. In this nanopore simulation, the setup using the fine mesh results in an equation system with $\num{367,431}$ unknowns, while the setup using the coarse mesh results in only $\num{26,895}$ unknowns, making the fine-mesh simulation at least a factor of $13.7$ times more expensive, possibly much more if a linearly scaling solver can not be used. For the combination (ii) a fine mesh and traditional fluid coupling~\cref{eq:model_forcedens_estatics} (green diamonds), the results are indeed physically sensible with a net-flow rate of zero for zero applied voltage bias. 

The third combination (iii) of a coarse grid and our improved coupling term (red circles) highlights the importance of our results. Our improved coupling force~\cref{eq:model_forcedens_friction} is able to reproduce the same physics as (ii) without the increase in computational cost that comes with increasing the grid resolution, as we predicted in Section~\ref{sec:corrected_coupling}. 

Finally, (iv) combining our improved coupling~\cref{eq:model_forcedens_friction} and a fine mesh (black triangles) reproduces the same physics as (ii) and (iii), but at the same increase in computational cost as (ii). We therefore conclude that using our improved coupling and the coarse grid (iii) is preferred on account that increasing the grid resolution using the improved coupling (iv) does not significantly improve accuracy of results as compared with a coarse grid and our improved coupling (iii), and at the same time reduces computational cost compared with a fine mesh and traditional coupling (ii). 

To summarise the methods we have introduced in this results section, here is a sample recipe one can follow to limit spurious flow in FEM simulations of electrokinetic phenomena, using a commercial FEM simulation package such as COMSOL: 

\begin{enumerate}
\item Setup the system geometry.
\item Specify the individual diffusion-advection, electrostatics, and hydrodynamic equations. 
\item Specify boundary conditions. 
\item Couple the individual equations to set-up the non-linear equation system.
\item When specifying the driving force in the hydrodynamic equations, choose a driving force term as presented in this paper $\vec{f} = -\textstyle\sum_{k} ( \kT \nabla c_{k} + z_{k} e c_{k} \nabla \Phi )$ rather than the traditional $\vec{f} = z_{k} e c_{k} \nabla \Phi$. 
\end{enumerate}

\section{Conclusions}
\label{sec:conclusion}

We have shown both theoretically and numerically that simulations of electrokinetic phenomena frequently suffer from spurious flow and spurious fluxes that can distort results and make their numerical treatment unnecessarily costly. We have also shown that previous approaches to suppress these artefacts by Capuani \textit{et al.}\cite{capuani_discrete_2004}\ produce correct solutions to the electrokinetic equations only in equilibrium and incur higher-order discretisation errors in non-equilibrium situations that grow exponentially with the local electric field and the grid size. While these errors remain small for typical simulations, such as for the electrophoresis of charged colloids, they can be significant in simulations of nanopores, where strong local electric fields exist. We demonstrated that a direct discretisation of the relevant equations eliminates these exponentially unbounded errors. 

Finally, we have proposed a method to limit spurious flow in numerical simulations of electrokinetic phenomena. Our method involves adding an additional gradient term to the fluid coupling in the electrokinetic equations. We demonstrated that this change does not affect the solutions for the fluid velocity, but does decrease spurious flow by several orders of magnitude using both a time-independent solver by Capuani \textit{et al.}\cite{capuani_discrete_2004}\ and a finite-element solver to simulate a charged sphere in an electrolyte solution. We verified the advantages of our improved coupling scheme with simulations of a nanopore, showing that using our improved coupling method with a coarse mesh produces the same results as using a fine mesh and the traditional coupling scheme used in the literature to date, but with an order of magnitude reduction in computational cost.  

Our results have particularly important implications for the numerical simulation of non-equilibrium phenomena such as electro-osmotic flow in nanopores. If simulations are carried out according to the commonly-used algorithms in the present literature, they may lead to inaccurate and unphysical results.

\section*{Acknowledgements}

GR, JdG, and CH thank the DFG for funding through the ``SPP 1726. Microswimmers  ---  From Single Particle Motion to Collective Behaviour''.  GR and CH acknowledge further funding from the SFB715, TPC.5, and. JdG gratefully acknowledges  financial  support  by  an NWO Rubicon Grant (\#680501210). We thank M.\ Kuron, O.\ A.\ Hickey, U. D. Schiller, and I.\ Pagonabarraga for useful discussions.

\bibliography{references}

\end{document}